\documentclass[a4paper,twocolumn,11pt]{quantumarticle}
\pdfoutput=1
\usepackage[utf8]{inputenc}
\usepackage[english]{babel}
\usepackage[T1]{fontenc}
\usepackage{amsmath}
\usepackage{hyperref}

\usepackage{tikz}
\usepackage{lipsum}

\usepackage{graphicx}
\usepackage[english]{babel}
\usepackage{amsmath}
\usepackage{float}
\usepackage{hyperref}
\usepackage{subfigure}
\usepackage{tikz}
\usepackage{physics}
\usepackage{breqn}
\usepackage{mathtools}
\usepackage{enumerate}
\usepackage{amssymb}
\usepackage{mathrsfs}
\usepackage{dsfont}
\usepackage{soul}

\usepackage[numbers]{natbib}

\def\ket#1{\vert{#1}\rangle}

\newcommand{\s}[1]{\mathrm{#1}}

\def\gc{\Gamma_{\s{C}}}
\def\gh{\Gamma_{\s{H}}}
\def\gmmc{\Gamma_{\s{C0}}^-}

\def\gmmh{\Gamma_{\s{H0}}^-}
\def\gmmh{\Gamma_{\s{H0}}^+}
\def\tc{T_{\s{C}}}
\def\th{T_{\s{H}}}
\def\r{\hat{\varrho}}
\def\dis#1{\mathcal{D}[#1]}

\begin{document}

\title{Steady-state entanglement production in a quantum thermal machine with continuous feedback control}

\author{Giovanni Francesco Diotallevi}
\affiliation{Physics Department and NanoLund$,$ Lund University$,$ Box 118$,$ 22100 Lund$,$ Sweden.}
\affiliation{Augsburg University$,$ Institute of Physics$,$ Universitätsstraße 1 (Physik Nord)$,$ 86159 Augsburg}
\email{francesco.diotallevi@uni-a.de}

\author{Bj\"orn Annby-Andersson}
\affiliation{Physics Department and NanoLund$,$ Lund University$,$ Box 118$,$ 22100 Lund$,$ Sweden.}

\author{Peter Samuelsson}
\affiliation{Physics Department and NanoLund$,$ Lund University$,$ Box 118$,$ 22100 Lund$,$ Sweden.}

\author{Armin Tavakoli}
\affiliation{Physics Department and NanoLund$,$ Lund University$,$ Box 118$,$ 22100 Lund$,$ Sweden.}

\author{Pharnam Bakhshinezhad}
\email{pharnam.bakhshinezhad@tuwien.ac.at}
\thanks{Formerly known as Faraj Bakhshinezhad.}
\affiliation{Physics Department and NanoLund$,$ Lund University$,$ Box 118$,$ 22100 Lund$,$ Sweden.}
\affiliation{Atominstitut, Technische Universit{\"a}t Wien, Stadionallee 2, 1020 Vienna, Austria.}

\maketitle

\begin{abstract}
 Quantum thermal machines can  generate steady-state entanglement by harvesting spontaneous interactions with local environments. However, using minimal resources and control, the entanglement is typically very noisy. Here, we study entanglement generation in a two-qubit quantum thermal machine in the presence of a continuous feedback protocol. Each qubit is measured continuously and the outcomes are used for real-time feedback to control the local system-environment interactions. We show that there exists an ideal operation regime where the quality of entanglement is significantly improved, to the extent that it can violate standard Bell inequalities and uphold quantum teleportation. In particular, we find, for ideal operation, that the heat current across the system is proportional to the entanglement concurrence. Finally, we investigate the robustness of entanglement production when the machine operates away from the ideal conditions.
\end{abstract}

\renewcommand{\thesection}{\Roman{section}}
\section{Introduction}

Quantum thermal machines are quantum systems coupled to two, or several, thermal reservoirs, which exploit temperature gradients to perform useful tasks such as cooling, heating, timekeeping, and producing work \cite{Kosloff-Ann.Rev.Phys.-2014,Mitchison-ContempPhys-2019,ThermodynamicsInTheQuantumRegime2019}. In contrast to their classical counterparts, these machines rely on quantum features, like entanglement and tunneling. Therefore, they are promising testbeds for studying fundamental aspects of quantum physics, such as the generation, stabilization, and control of entanglement in the presence of thermal environments.

To this end, it was shown that a minimal quantum thermal machine, consisting of two coherently interacting qubits coupled to two reservoirs at different temperatures, is able to produce stationary entangled states  \cite{Bohr_Brask_2015}. The word `minimal' refers to the minimal setup required to generate entanglement. The success of this machine can be linked to the magnitude of the heat current flowing through the system \cite{Shishir_heat_bound}.
However, the entanglement generated in such a machine is typically weak and noisy. For example, it is unable to perform well-known entanglement-based tasks such as teleportation or Bell inequality violation \cite{Brask-Quantum-2022}. Therefore, in order to improve the entanglement production, it has been considered to supply the original autonomous system with some additional resources. It has been found that heralding the output state of a multi-dimensional autonomous quantum thermal machine, via a local measurement, can generate maximally entangled states \cite{Tavakoli-Quantum-2018}. This type of approach also enables multipartite entanglement production \cite{PhysRevA.101.012315}. However, this requires coherent control of multi-level systems and the ability to perform non-demolition filter measurements. An alternative approach is to introduce a third bath that is common to both qubits \cite{Man-Phys.Scrip.-2019}, which leads to an improvement in the entanglement production. Another approach that improves the entanglement is to perform a population inversion process in fermionic baths  \cite{Brask-Quantum-2022}. This amounts to bath engineering, but can improve entanglement production to the extent that non-trivial teleportation fidelities are possible. Complementary to that, by implementing the minimal machine in a double quantum dot, a large voltage bias can be applied across the system to generate  entanglement that is nonlocal \cite{Prech-PRR-2023}.

In this paper, we investigate how the entanglement of the quantum thermal machine of Ref.~[4] can be controlled and improved using measurement-based feedback control. This route is independent of the experimental platform. We note that the idea of using measurement-based feedback to increase entanglement is not new, but has previously been explored in optical systems \cite{Wang-PRA-2005,Mancini-EurPhys-2005,Carvalho-PRA-2007,Carvalho-PRA-2008,Li-PRA-2008,Hou-PRA-2010}, where qubits enclosed in optical cavities are externally driven via feedback control. However, the lack of thermal environments and time-independent coherent interactions in he cited literature is a clear distinction from our setup. Our feedback protocol is based on a continuous parity measurement of the qubits, distinguishing whether one of the qubits are excited or if both are in the ground or excited state. If only one qubit is excited, the warmer bath is decoupled from the system, favoring coherent interactions between the qubits. If no or two excitations reside in the system, the hot bath is re-coupled to the qubits. The protocol is modeled by employing the quantum Fokker-Planck master equation presented in Ref.~\cite{Annby-PRL-2022}, which was developed to describe continuous, Markovian feedback protocols like the one presented here. To facilitate a direct comparison with previous relevant works, we use the concurrence to quantify the entanglement, as well as investigate operational aspects of the nonclassicality of the produced entanglement \cite{Brask-Quantum-2022}. We identify an optimal operation regime where the concurrence significantly exceeds what was found in the elementary machine Ref.~\cite{Bohr_Brask_2015}. Additionally, we find that the entanglement can violate the CHSH inequality and uphold quantum teleportation. In particular, we find, in the optimal regime, that the concurrence is proportional to the heat current flowing between the reservoirs, implying that a nonzero heat current is an entanglement witness. This contrasts the elementary machine, in which the heat current must exceed a non-trivial threshold to act as an entanglement witness \cite{Shishir_heat_bound}. We note that the results obtained in the optimal operation regime are independent whether the thermal reservoirs are bosonic or fermionic. We also investigate the robustness of the entanglement production when relaxing the ideal conditions. We find that the entanglement decreases, while still being larger than in the absence of feedback.

The paper is structured as follows. In Sec.~\ref{sec: system}, we briefly review the system and the concurrence as a measure for entanglement. Section \ref{sec: feedback protocol} introduces the feedback protocol and how it is modeled. In Sec.~\ref{sec: results}, we present our results, and Sec.~\ref{sec:conclusion} concludes the paper.

\section{System and entanglement} \label{sec: system}
\begin{figure}[H]
\centering
\includegraphics[width=.8\linewidth]{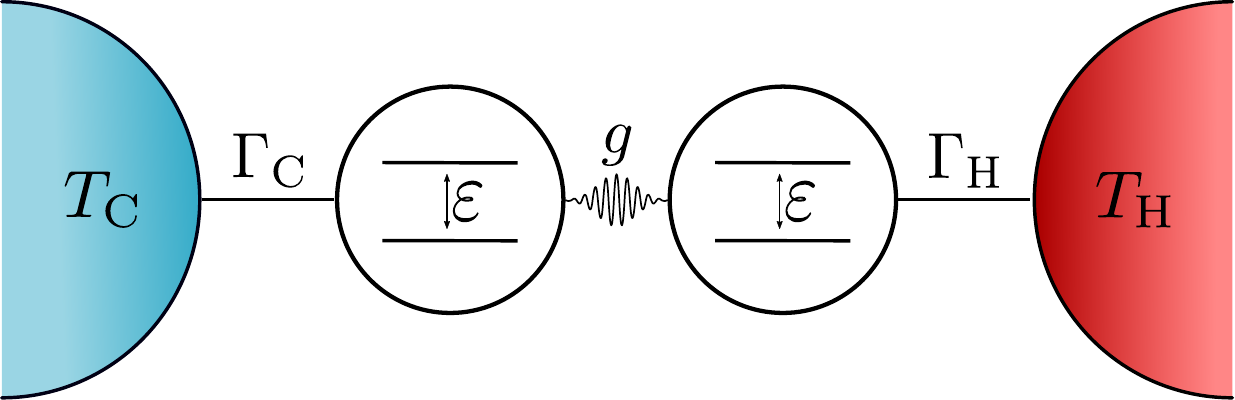}
\caption{A sketch of the system. Two coherently interacting qubits are coupled, via the bare tunnel rates $\Gamma_{\rm C}$ and $\Gamma_{\rm H}$, to two separate thermal reservoirs of different temperatures ($T_{\rm C}<T_{\rm H}$). The strength of the coherent interaction is parametrized by $g$.}
\label{fig: system sketch1}
\end{figure}

We consider two coherently interacting qubits coupled to two thermal reservoirs with temperatures $T_{\rm C}$ and $T_{\rm H}$ ($T_{\rm C}<T_{\rm H}$) as depicted in Fig.~\ref{fig: system sketch1}. Note that the coherent interaction is autonomous, and does not rely on external driving. The reservoirs can be fermionic or bosonic, but as many of the results are independent of particle type, we introduce the system without specifying particle type, keeping the discussion general. If particle type matters, we will clearly specify this. We consider the following Hamiltonian of the qubits,
\begin{align}
\hat{H} = \varepsilon &\left( \dyad{1}_{\s{C}}\otimes\mathds{1} + \mathds{1}\otimes\dyad{1}_{\s{H}} \right) \label{eq: general hamiltonian} \\ \nonumber &+ g\left( \dyad{01}{10}+\dyad{10}{01} \right) + U\dyad{11}, 
\end{align}
where $\ket{0}$ and $\ket{1}$ denote the ground and excited states of the qubits, $\mathds{1}$ is the identity operator, $\varepsilon$ is the energy of the excited state of each qubit, $g$ is the strength of the coherent flip-flop interaction, and $U$ is the interaction energy between the excited states. The last term naturally arises when the excitations carry charge. 

By assuming weak system-bath and qubit-qubit interactions, the dynamics of the system can be written as a local Lindblad master equation (we set $\hbar=1$),
\begin{align}
   \label{eq: Free evolv lindblad master eqn}
   \partial_t \r= &\mathcal{L}\r = i[\r , \hat{H}] \\
    \nonumber &+ \sum_{\substack{k\in\{\rm C,H\}\\ l\in\{0,1\}}} \left(\Gamma_{kl}^+  \dis{\hat{J}^\dagger_{kl}} + \Gamma_{kl} ^-\dis{\hat{J}_{kl}}\right) \r, 
\end{align}
where we introduced the shorthand superoperator notation $\mathcal{L}$, describing the dynamics of the system. The dissipators $\dis{\hat{J}_{kl}}\r \equiv \hat{J}_{kl} \varrho \hat{J}_{kl}^\dagger - \frac{1}{2} \left\{ \hat{J}_{kl}^\dagger \hat{J}_{kl}, \r\right\}$, where $\hat{J}_{\rm C0}=\dyad{00}{10}$, $\hat{J}_{\rm 
 C1}=\dyad{01}{11}$, $\hat{J}_{\rm 
 H0}=\dyad{00}{01}$, and $\hat{J}_{\rm 
 H1}=\dyad{10}{11}$ are jump operators describing bath-induced de-excitations of the qubits with corresponding excitation (+) and de-excitation $(-)$ rates
\begin{align}
\label{eq: rates}
    \Gamma_{kl} ^+ = \frac{\Gamma_k}{e^{(\varepsilon+lU)/T_k}\pm 1}, \hspace{0.5cm} \Gamma_{kl} ^- = \frac{\Gamma_k}{1\pm e^{-(\varepsilon+lU)/T_k}},
\end{align}
where $\Gamma_k$ is the bare transition rate for bath $k$ (see Fig.~\ref{fig: system sketch1}), and the $+$ ($-$) in the denominators corresponds to fermionic (bosonic) reservoirs. Note that the chemical potentials of the baths are set to zero. From here on, we normalize all energies with respect to $\varepsilon$, i.e., effectively setting $\varepsilon=1$.

The stationary state of Eq.~(\ref{eq: Free evolv lindblad master eqn}) takes the form
\begin{align}
        \r_\infty = \begin{bmatrix}
\varrho_{00} &0  & 0 &0 \\ 
 0& \varrho_{01} & \alpha &0 \\ 
 0& \alpha^* & \varrho_{10} &0 \\ 
 0& 0 & 0 & \varrho_{11}
\end{bmatrix}, 
\label{eq: simple steady form}
\end{align}
when written in the computational basis $\{\ket{00},\ket{01},\ket{10},\ket{11}\}$. Note that $\sum_{\s{i,j}=0}^1\varrho_{\s{ij}}=1$ and $|\alpha|\hspace{0.1cm}\leq\sqrt{\varrho_{01}\varrho_{10}}$ ensure the normalization and positivity of $\r$. This form in Eq.\,(\ref{eq: simple steady form}) arises because of the flip-flop interaction in the Hamiltonian in Eq.\,(\ref{eq: general hamiltonian}) and the dissipative interactions between the system and the reservoirs, only allowing coherent interaction in the subspace $\{\ket{01},\ket{10}\}$ such that the remaining coherences vanish for long times. To quantify the entanglement in the stationary state, we use the concurrence~\cite{Concurrence_wootters_1998}, which is an entanglement monotone for bipartite systems that can identify fully separable and maximally entangled states. For the state in Eq.~(\ref{eq: simple steady form}), the concurrence takes the form \cite{Brask-Quantum-2022}
\begin{align}
    \mathcal{C}(\r_\infty) = \s{max}\{2(|\alpha| - \sqrt{\varrho_{00}\varrho_{11}}), 0\}, 
    \label{eq: simple concurrence}
\end{align}
taking values between 0 and 1, where 0 corresponds to $\r_\infty$ being fully separable and 1 to being maximally entangled. Values different from 0 and 1 correspond to states that are not maximally entangled. In Ref.~\cite{Bohr_Brask_2015} it was shown that the maximal stationary concurrence is given by $\mathcal{C}\sim0.09\,(0.25)$ for bosonic (fermionic) particles. The difference in performance is due to $U\neq0$ for fermions, which is naturally the case for, e.g., electrons. However, the entanglement generated this way is not useful in several operational notions of nonclassicality, such as steering, nonlocality and teleportation \cite{Brask-Quantum-2022}.

\section{Feedback Protocol} \label{sec: feedback protocol}
\begin{figure*}[tb] 
\centering
 \includegraphics[scale=0.4]{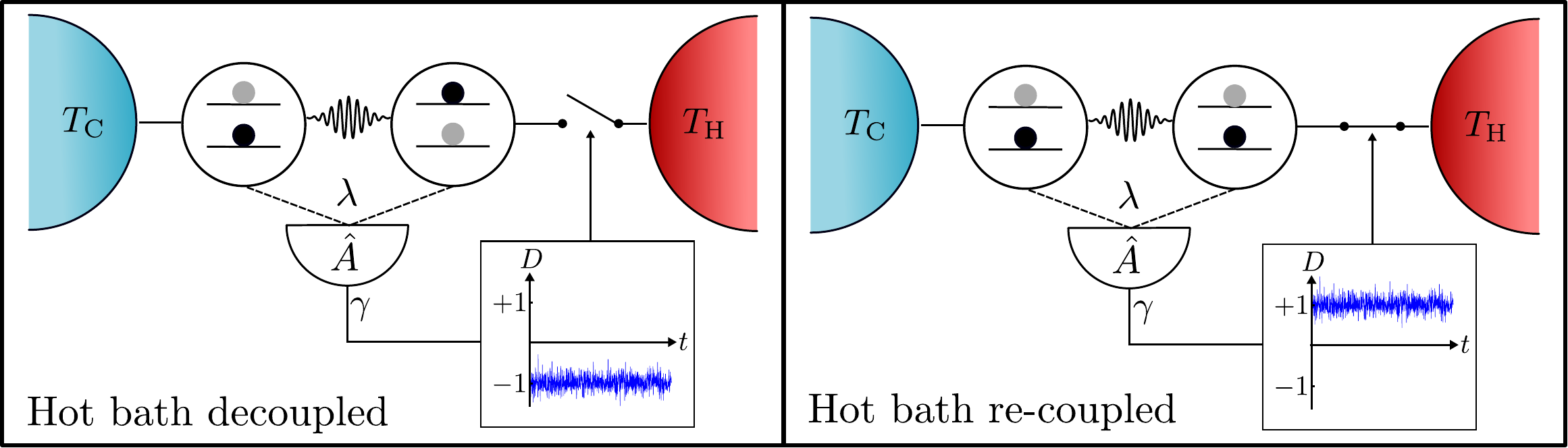}
\caption{A qualitative sketch of the feedback protocol. A detector with finite bandwidth $\gamma$ is coupled to the system via measurement strength $\lambda$, and performs continuous measurements of the observable $\hat{A}=\hat{\sigma}_z\otimes\hat{\sigma}_z$. Based on the detector outcome $D$, the coupling with the hot reservoir is switched on or off dependent on whether one excitation or no, or two, excitations reside in the system. The black and grey dots illustrate the various excitation configurations.}
\label{fig: feedback sketch}
\end{figure*}

To increase the entanglement, it necessary to transfer population from the $\{\ket{00},\ket{11}\}$ subspace to the $\{\ket{01},\ket{10}\}$ subspace, while simultaneously maximizing the coherence $\alpha$ in the latter subspace, see Eqs.~(\ref{eq: simple steady form}) and (\ref{eq: simple concurrence}). To this end, we introduce a feedback protocol, controlling the coupling between the system and the hot reservoir. The control procedure follows an on-off protocol~\cite{Astrom-book} and is conditioned on measuring the parity of the qubits, see the qualitative sketch in Fig.~\ref{fig: feedback sketch}. Note that when measuring parity, it is not possible to distinguish which of the qubits that is excited. Therefore, the measurement does not affect the coherence $\alpha$. When a single excitation resides in the system, the hot bath is decoupled from the hot qubit. An excitation in the hot qubit can thus only interact with the cold qubit, and not dissipate into the hot bath. This reduces $\varrho_{00}$ and $\varrho_{11}$, while increasing the coherently coupled populations $\varrho_{01}$ and $\varrho_{10}$. Also note that the decoherence induced by the hot bath is suppressed, shielding the coherence of the system. With no, or two, excitations in the system, the hot bath is re-coupled, again allowing thermal excitation of the hot qubit. Note that it is not useful to additionally close the coupling to the cold reservoir, as excitations would oscillate between the qubits indefinitely, preventing stationary entanglement production. 

Mathematically, we formulate the feedback protocol in the following way. We continuously measure the parity observable
\begin{align}
    \hat{A} = \hat{\sigma}_z\otimes\hat{\sigma}_z, \label{eq: parity observable}
\end{align}
where $\hat{\sigma}_z$ is the Pauli-Z matrix. Note that $[\hat{A},\r_\infty]=0$ for the density matrix in Eq.~(\ref{eq: simple steady form}), resulting in a backaction-free measurement \cite{Annby-PRL-2022}. This means that the coherence in Eq.~(\ref{eq: simple steady form}) will not be affected by the measurement. As a result, the measurement is not detrimental for the entanglement production. The detector output $D$ is noisy, with fluctuations around $-1$ when the system occupies $\ket{01}$ or $\ket{10}$, and around $+1$ when occupying $\ket{00}$ or $\ket{11}$, see the time traces in Fig.~\ref{fig: feedback sketch}. We thus interpret the signal as follows. When $D<0$, we assume that $\ket{01}$ or $\ket{10}$ is occupied. For $D>0$, we assume that $\ket{00}$ or $\ket{11}$ is occupied.



To describe the dynamics of the system under the feedback protocol, we make use of the quantum Fokker-Planck master equation introduced in Ref.~\cite{Annby-PRL-2022}. This formalism allows us to describe the dynamics of any quantum system undergoing continuous, Markovian feedback control. For our protocol, it reads
\begin{align}
     \partial_t \r_t(D)  &= \mathcal{L}(D)\r_t(D) \label{eq: Fokker-Planck Master Equation} \\
     \nonumber &- \gamma \partial_D \mathcal{A}(D)\r_t(D) + \frac{\gamma^2}{8\lambda}\partial_D^2\r_t(D),
\end{align}
where $\r_t(D)$ is the joint system-detector state, with $\r_t=\int dD\r_t(D)$ being the system state independent of the detector, and $p_t(D)=\trace\{\r_t(D)\}$ being the probability distribution of observing outcome $D$ at time $t$. Note that $\int dD\trace\{\r_t(D)\}=1$.

The feedback-controlled dynamics of the system are described by 
\begin{align}
    \mathcal{L}(D) = \theta(D)\mathcal{L} + [1-\theta(D)]\tilde{\mathcal{L}}, \label{eq: feedback operation}
\end{align}
where $\mathcal{L}$ is given by Eq.~(\ref{eq: Free evolv lindblad master eqn}) and describes the dynamics when the hot bath is coupled to the system, while $\tilde{\mathcal{L}}\r=i[\r, \hat{H}] +  \sum_{l\in\{0,1\}} \left(\Gamma_{\text{C} l}^+  \dis{\hat{J}^\dagger_{\text{C}l}} + \Gamma_{\text{C}l} ^-\dis{\hat{J}_{\text{C}l}}\right) \r$ describes the dynamics when decoupling the hot bath.

The remaining two terms of Eq.~(\ref{eq: Fokker-Planck Master Equation}) constitute a Fokker-Planck equation describing the time evolution of the detector. The superoperator drift coefficient $\mathcal{A}(D)\r \equiv \frac{1}{2}\{\hat{A}-D, \r \}$ describes the coupling between the system and detector, and determines the average position of the detector, dependent on the system state. Note that $\gamma$ is the bandwidth of the detector, such that $1/\gamma$ gives the lag of the detector. The last term describes the diffusion of the detector position, where the diffusion constant $\gamma/8\lambda$ corresponds to the noise of the detector. Here $\lambda$ is the strength of the measurement. The limit $\lambda \to 0$ corresponds to a weak measurement. In this limit, the noise increases, and thus also the uncertainty of the measurement. The limit $\lambda \to \infty$ corresponds to a projective measurement, where the noise vanishes, eliminating all uncertainty.

Here we focus on the regime $\gamma\gg\text{max}\{g,\Gamma_{kl}^\pm\}$, where the detector is much faster than the dynamics of the system. This is beneficial for entanglement production as the detector never lags behind the system, reducing feedback mistakes due to detector delay. However, as the ratio between $\gamma$ and $\lambda$ determines the magnitude of the noise, feedback mistakes due to noise can still occur. For a fast detector ($\gamma\gg\text{max}\{g,\Gamma_{kl}^\pm\}$), Eq.~(\ref{eq: Fokker-Planck Master Equation}) can be reduced to a Markovian master equation for the system alone \cite{Annby-PRL-2022}. It is given by
\begin{align}
    \label{eq:MEIdealMeas}
    \partial_t \r_t =& \mathcal{L}_{\rm fb} \r_t,
\end{align}
where the feedback-controlled dynamics are described by $\mathcal{L}_{\rm fb}$.  In Appendix\,\,\ref{appendix: separation of time scales}, we detail the derivation of this equation, where we also give the general form of $\mathcal{L}_{\rm fb}$.  In Sec.~\ref{sec:Ideal operation}, where we study ideal conditions, we present a simple representation of $\mathcal{L}_{\rm fb}$.

\section{Results} \label{sec: results}

\subsection{Ideal operation}
\label{sec:Ideal operation}
\begin{figure*}[tb] 
\centering
 \makebox[\textwidth]{\includegraphics[width=0.8\paperwidth]{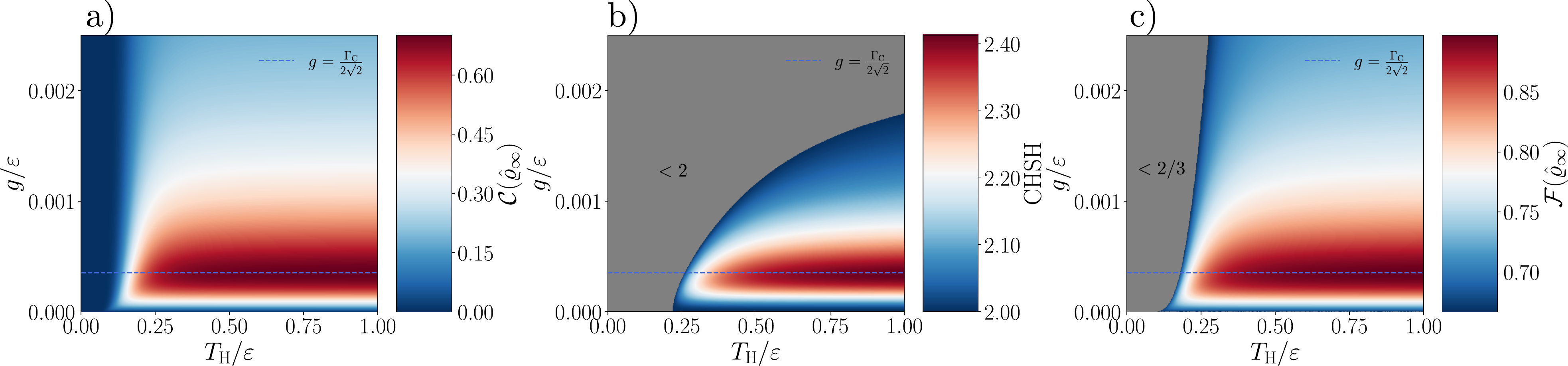}}
\caption{Steady state results for the concurrence (a), $\s{CHSH}$ (b) and teleporation fidelity (c) as a function of $g$ and $\th$. These plots were obtained by numerically solving Eq.~(\ref{eq:MEIdealMeas}), using close-to-ideal conditions, see specification of parameters at the end of the caption. The blue dashed line in each plot marks the optimal value of $g$ for entanglement generation. The grey, shaded areas in (b) and (c) mark where the CHSH and transportation fildelity are below the classical limit. The three graphs were obtained considering $\gc/\varepsilon = 10^{-3}$, $\gh = 100\gc;$  $\lambda/ \varepsilon = 100$; $U/ \varepsilon = 100$; $\gamma/\varepsilon = 1$; $\tc/\varepsilon = 10^{-2} $.}
\label{fig: optimal cond. results}
\end{figure*}

To optimize the entanglement generation in the system, it is useful to study limiting cases of various parameters. Here we study two such limits, and refer to these as ideal operation conditions -- this is motivated in Figs.~\ref{fig: finite lambda different gh} and \ref{fig: finite lambda fermionic} below. The first limit is $T_{\rm C}\to0$, ensuring that the cold bath cannot thermally excite the cold qubit ($\Gamma_{\text{C}l}^+\to0$), and thus reduces the population of $\ket{11}$. The second limit is $\lambda\to\infty$, which is physically motivated as the measurement is backaction-free. Therefore, this limit is not detrimental for the entanglement generation. Additionally, this limit completely suppresses the noise in the detector, such that feedback is always applied correctly.

Under these limits, the Liouville superoperator in Eq.~(\ref{eq:MEIdealMeas}) can, by vectorizing the nonzero elements of the density matrix as $\r=(\varrho_{00}, \varrho_{01}, \varrho_{10}, \varrho_{11}, \alpha, \alpha^*)^T$, be written in matrix representation as
\begin{widetext}
    \begin{align}
    \mathcal{L}_{\rm fb} =\left[\begin{matrix}
 - \Gamma_{\s{H}0}^+  & 0 & \Gamma_{\s{C}0}^- & 0 & 0 & 0\\
\Gamma_{\s{H}0}^+  & 0& 0 & \Gamma_{\s{C}1}^- & i g & - i g\\
0 & 0 &  - \Gamma_{\s{C}0}^- & \Gamma_{\s{H}1}^-  & - i g & i g\\
0 & 0 & 0 & - \Gamma_{\s{C}1}^- - \Gamma_{\s{H}1}^-  & 0 & 0\\
0 & i g & - i g & 0 & -\frac{\Gamma_{\s{C}0}^-}{2} & 0\\
0 & - i g & i g & 0 & 0 & -\frac{\Gamma_{\s{C}0}^-}{2} 
\end{matrix}\right].
\label{eq: Liouvillian optimal conditions}
\end{align}
\end{widetext}
Under ideal operation conditions, excitations are unidirectionally transported from the hot to cold reservoir. This is beneficial for entanglement production as an excitation in the hot qubit only can interact coherently with the cold qubit, thus reducing the population of $\ket{00}$. Additionally, we note that the doubly excited state $\ket{11}$ is decoupled from the remaining states, and will thus vanish in the stationary state.

The null-space of $\mathcal{L}_{\rm fb}$ corresponds to the stationary state of Eq.~(\ref{eq:MEIdealMeas}) and provides the following stationary concurrence of the system (see Appendix\,\,\ref{appendix: steady state calculations})
\begin{align}
    \mathcal{C}(\r_\infty) = \frac{4g \Gamma_{\text{C}0}^- \Gamma_{\text{H}0}^+}{4g^2 \left(\Gamma_{\text{C}0}^-+2 \Gamma_{\text{H}0}^+\right)+ \left(\Gamma_{\text{C}0}^-\right)^2 \Gamma_{\text{H}0}^+}. \label{eq: ss concurrence}
\end{align}
The concurrence attains its maximum $\mathcal{C}(\r_\infty)=1/\sqrt{2}\approx0.71$ when $\Gamma_{\text{H}0}^+\gg\Gamma_{\text{C}0}^-$ and $\Gamma_{\text{C}0}^-/g=2\sqrt{2}$, thus significantly increasing the concurrence obtained in the absence of measurement and feedback \cite{Bohr_Brask_2015}. The condition $\Gamma_{\text{H}0}^+\gg\Gamma_{\text{C}0}^-$ ensures that an excitation quickly enters the system via the hot bath when the system occupies $\ket{00}$. This increases the population in the subspace $\{\ket{01},\ket{10}\}$, favoring entanglement generation. Note that increasing $g$ indefinitely is detrimental for the entanglement production, as it enhances Rabi oscillations in the coherently coupled subspace. Averaging over many oscillations reduces the entanglement [see Eq.\,(\ref{eq: ss concurrence})]. 

We note that the same concurrence (\ref{eq: ss concurrence}) was obtained in Ref.~\cite{Prech-PRR-2023} when implementing the system in a double quantum dot with $U\to\infty$ and an infinite external voltage bias across the system. Under these conditions, the system autonomously evolves according to Eq.~(\ref{eq: Liouvillian optimal conditions}), even in the absence of measurement and feedback. However, we stress that our results also are valid for non-interacting particles ($U=0$).

As excitations are transported from the hot to cold bath, heat will flow through the system. Due to the coherent interaction between the qubits, a nonzero heat current indicates the presence of coherence in the system, and is necessary for entanglement production \cite{Bohr_Brask_2015}. Under ideal operation conditions, the heat current is given by (derivation in Appendix \ref{app:heat exchange})
\begin{align}
    \dot{Q} = -\Gamma_{\text{C}0}^- \trace\{\hat{H}\mathcal{D}[\hat{J}_{\text{C}0}]\r_\infty\} = \varepsilon g \mathcal{C}(\r_\infty),
    \label{eq: concurrence as funct of heat}
\end{align}
with $\r_\infty=\int dD\r_\infty(D)$. We stress that Eq.~(\ref{eq: concurrence as funct of heat}) is independent of particle type. The relation implies that the concurrence can be directly inferred by measuring the heat current, and does not require quantum state tomography. This implies that the heat current is an entanglement witness, where a nonzero current indicate the presence of entanglement. 

While the concurrence indicates whether a state is entangled or not, it does not provide any information on how useful the entanglement is for quantum information processing. Therefore, it is useful, as a complement to the concurrence, to evaluate if an entangled state is able to perform useful tasks in quantum information processing. To this end, we evaluate whether the generated entanglement can violate the CHSH inequality and perform quantum teleportation \cite{Brask-Quantum-2022}. 

For the state in Eq.~(\ref{eq: simple steady form}), the CHSH inequality may be expressed as $\text{CHSH}\leq2$, where \cite{Brask-Quantum-2022,Prech-PRR-2023}
\begin{align}
\label{eq: CHSH}
    &\s{CHSH} =\\
    &\,\,\,2\sqrt{8\alpha^2 + (2\Delta - 1)^2 - \s{min}\{4\alpha^2, (2\Delta - 1)^2\}}, \nonumber 
\end{align}
with $\Delta \equiv \varrho_{01} + \varrho_{10}$. If $\text{CHSH}>2$, the system state shows Bell nonlocality. For maximally entangled states $\text{CHSH}=2\sqrt{2}$. At maximum concurrence for ideal operation conditions, $\text{CHSH}=\sqrt{6}\approx2.45$. In fact, this is the maximum value the CHSH can attain under the feedback protocol, see Appendix \ref{appendix:CHSH}.

To quantify how well $\r_\infty$ can perform quantum teleportation, we calculate the teleportation fidelity $\mathcal{F}(\r_\infty) = [1+2F(\r_\infty)]/3$, where $F(\r_\infty)$ is the singlet fraction for a two qubit system $\r_\infty$ (see Appendix~\ref{appendix: singlet fraction})~\cite{Horodeki_1999_sfract}. For states taking the form of Eq.~(\ref{eq: simple steady form}), the singlet fraction is expressed as~\cite{Brask-Quantum-2022}
\begin{align}
 F(\r)\,=\,
\begin{cases}
\alpha + \frac{\Delta}{2} &\hspace{-1cm}\s{if} \,(1+2\alpha - 2\Delta) \leq 0 \\[0.5mm]
\s{max}\{\alpha + \frac{\Delta}{2}, \frac{1-\Delta}{2}\}  &\s{otherwise}
\end{cases}, \label{eq: Singlet Fraction}
\end{align}
\vspace{-0.5cm}

For a maximally entangled state, $\mathcal{F}(\r_\infty)=1$. A classical implementation of the protocol can at best achieve $\mathcal{F} = 2/3$~\cite{Horodeki_1999_sfract}, implying that the state contains useful entanglement when $F(\r_\infty) > 1/2$. At maximum concurrence, for ideal operation conditions, we obtain $\mathcal{F}(\r_\infty)=(4+\sqrt{2})/6\approx0.9$. Note that this is the maximum fidelity that can be achieved with the feedback protocol, see Appendix \ref{appendix: singlet fraction}. In Fig.~\ref{fig: optimal cond. results}, we plot the concurrence, CHSH, and teleportation fidelity, using close-to-ideal conditons.


\subsection{Beyond ideal operation} \label{beyond ideal operation}

\begin{figure*}[tb] 
\centering
 \makebox[\textwidth]{\includegraphics[width=.8\paperwidth]{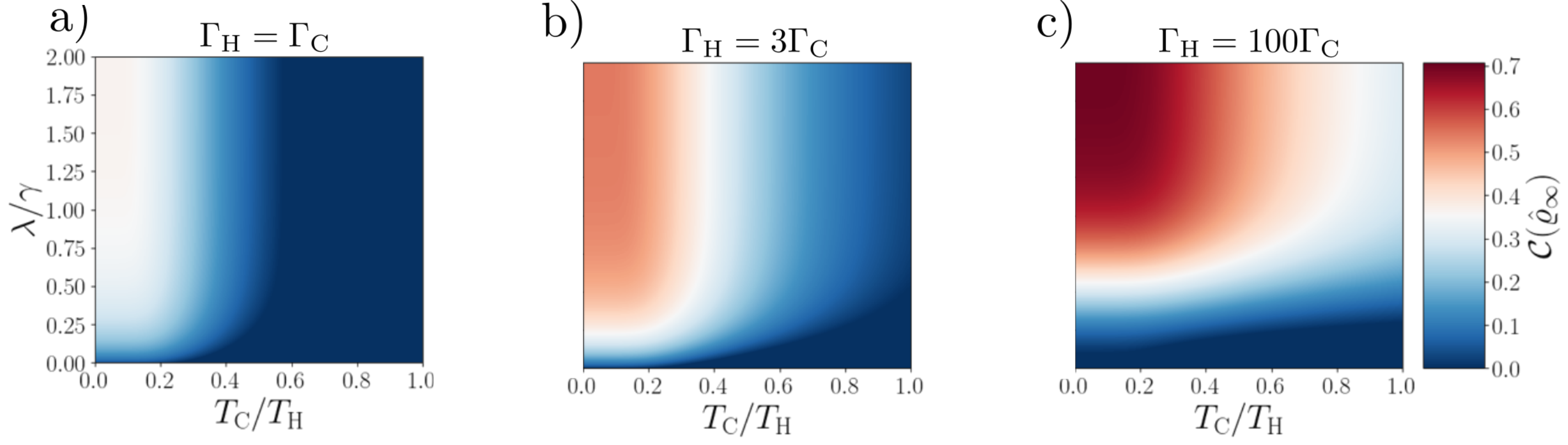}}
\caption{Steady state concurrence as a function of measurement strength $\lambda$ and temperature of the cold reservoir $\tc$ for three values of $\Gamma_{\rm H}$ (see top of graphs). Here we focus on fermionic reservoirs (similar results are obtained for bosons, see Appendix \ref{appendix:concurrence,CHSH,fidelity}). All plots indicate that the concurrence decreases with the measurement strength $\lambda$, as feedback mistakes become prominent. By increasing $T_{\rm C}$, the cold qubit can be thermally excited, increasing the population of $\ket{11}$ and thus reducing the concurrence. For $\Gamma_{\rm H}\gg\Gamma_{\rm C}$, the hot bath quickly provides an excitation to the system when occupying $\ket{00}$, reducing the population of this state and favors entanglement generation. All of the above plots were obtained considering $U/\varepsilon = 0$, $\gc/\varepsilon = 10^{-3}$,  $\gc/g = 2\sqrt{2}$ and $\th/\varepsilon =1 $, $\gamma/\varepsilon = 1 $.}
\label{fig: finite lambda different gh}
\end{figure*}

\begin{figure*}[tb] 
\centering
 \makebox[\textwidth]{\includegraphics[width=.8\paperwidth]{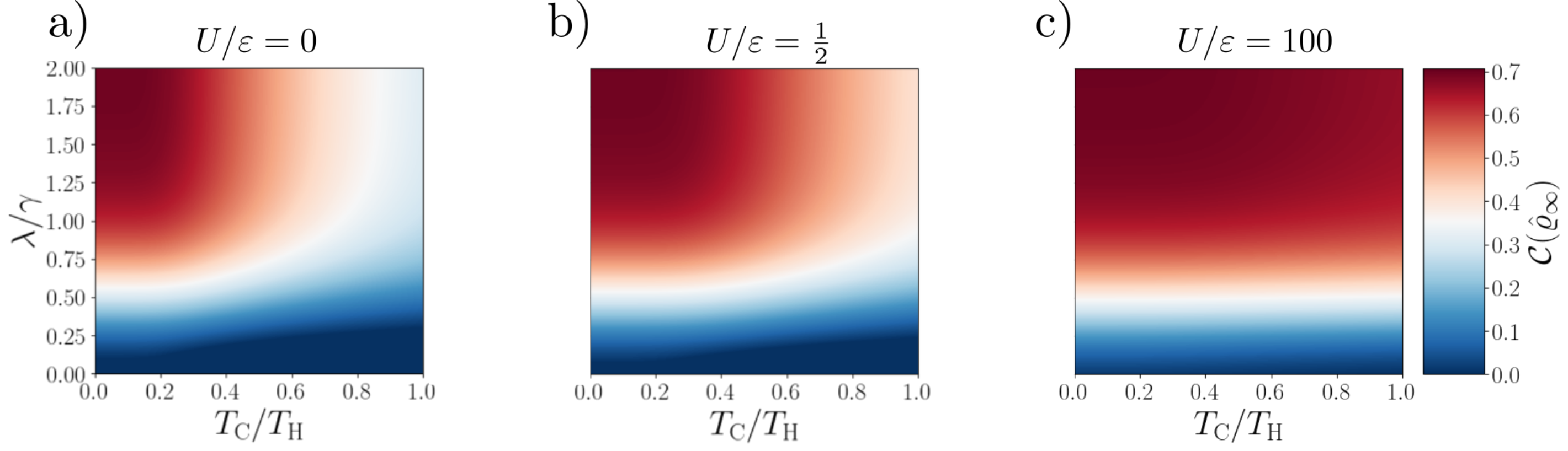}}
\caption{Steady state concurrence as a function of measurement strength lambda $\lambda$ and the temperature of the cold bath $\tc$ for three values of $U$. Here we focus on fermionic reservoirs (similar results are obtained for bosons, see Appendix \ref{appendix:concurrence,CHSH,fidelity}). The concurrence shows a similar behavior as in Fig.~\ref{fig: finite lambda different gh}. When increasing $U$, the concurrence becomes invariant of $\tc$, as the population of $\ket{11}$ vanishes, thus preventing thermal excitations of the cold qubit if the hot one is excited. The graphs were obtained using $\gc/\varepsilon = 10^{-3}$, $\gh = 100\gc;$  $\gc/g = 2\sqrt{2}$ and $\th/\varepsilon = 1$, and $\gamma/\varepsilon = 1$. }
\label{fig: finite lambda fermionic}
\end{figure*}


Now we discuss the entanglement production when relaxing the ideal operation conditions. The ideal conditions are relaxed one by one, such that the role of each parameter can be understood carefully. We also discuss the effect of the last term of the Hamiltonian in Eq.\,(\ref{eq: general hamiltonian}). We focus on investigating how the concurrence depends on the system parameters. For a similar analysis of CHSH and the quantum teleportation fidelity, the reader is referred to Appendix \ref{appendix:concurrence,CHSH,fidelity}. However, such an analysis does not provide any additional information compared to the concurrence. The figures presented in this section focus on fermionic reservoirs [see Eq.~(\ref{eq: rates})], but a similar behavior is observed for bosonic reservoirs, see Appendix \ref{appendix:concurrence,CHSH,fidelity}.

By relaxing $\lambda\to\infty$, the detector becomes noisy -- recall that the magnitude of the noise is determined by $\lambda/\gamma$, as discussed under Eq.~(\ref{eq: Fokker-Planck Master Equation}). A noisy detector introduces feedback mistakes. That is, the coupling to the hot bath can remain open even though $\ket{01}$ or $\ket{10}$ are occupied. This increases the populations of $\ket{00}$ and $\ket{11}$, because an excitation in the hot qubit can re-enter the hot bath when $\ket{01}$ is occupied, or enter the hot qubit when $\ket{10}$ is occupied. This reduces the entanglement in the system, see Figs.~\ref{fig: finite lambda different gh} and \ref{fig: finite lambda fermionic}. Additionally, the figures illustrate that it is favorable to use $\lambda\gg\gamma$, as seen in the previous subsection. We also note that feedback mistakes increase the decoherence induced by the hot bath, as the coupling to the hot bath can remain open even though $\ket{01}$ or $\ket{10}$ are occupied, see Eq.~(\ref{eq:full liouvillian}) in Appendix \ref{appendix: steady state calculations}. We remind the reader that the measured observable is backaction-free, such that the strength of the measurement does not affect the entanglement of the system.

For $T_{\rm C}\neq0$, $\Gamma_{\text{C}l}^+\neq0$, enabling thermal excitations of the cold qubit. The population of $\ket{11}$ thus increases as the cold qubit may be excited when occupying $\ket{01}$, decreasing the entanglement. We also note that $\Gamma_{\text{C}1}^+\neq0$ results in bath-induced decoherence, deteriorating the entanglement, see Eq.~(\ref{eq:full liouvillian}) in Appendix \ref{appendix: steady state calculations}. The overall effect of $T_{\rm C}\neq0$ is thus to decrease the entanglement -- this is illustrated in Figs.~\ref{fig: finite lambda different gh} and \ref{fig: finite lambda fermionic}.

Below Eq.~(\ref{eq: ss concurrence}), we noted that $\Gamma_{\text{H}0}^+\gg\Gamma_{\text{C}0}^-$ was favorable for entanglement generation as the population in the subspace $\{\ket{01},\ket{10}\}$ was increased. Relaxing this condition increases the population of $\ket{00}$, as it takes longer time for an excitation to enter the hot qubit when the system occupies $\ket{00}$. Similarly, when $\Gamma_{\rm C}^{-}$ is small, the population of $\ket{00}$ decreases, as an excitation stays longer in the system, favoring coherent interaction between the qubits. Thus, relaxing the condition $\Gamma_{\text{H}0}^+\gg\Gamma_{\text{C}0}^-$ decreases the entanglement as illustrated in Fig.~\ref{fig: finite lambda different gh}.

So far, we have not made any assumptions about the interaction $U$ in the Hamiltonian (\ref{eq: general hamiltonian}). In fact, under ideal operation conditions ($T_{\rm C}=0$ and $\lambda\to\infty$), the interaction does not play any role for the stationary state of the system as the population of $\ket{11}$ vanishes, see Eq.~(\ref{eq: Liouvillian optimal conditions}). However, when relaxing one, or both, of the ideal conditions, $U$ affects the stationary state as the population of $\ket{11}$ becomes nonzero. In Fig.~\ref{fig: finite lambda fermionic}, we illustrate the effect of $U$ on the concurrence. We see that the concurrence is dependent on $T_{\rm C}$ for non-interacting excitations ($U=0$), while a large $U$ eliminates this dependence. This happens because the population of $\ket{11}$ vanishes for large $U$, preventing thermal excitations when $\ket{01}$ is occupied.

For ideal conditions, we found that the heat current is proportional to the concurrence. In fact, this proportionality holds true when relaxing the ideal conditions, but taking the limit $U\to\infty$, where the population of $\ket{11}$ vanishes, see Appendix \ref{appendix: steady state calculations}. For finite $U$, this proportionality does not hold true anymore. This agrees with the results derived in Ref.~\cite{Shishir_heat_bound}, where it was found that the qubits, in the absence of feedback, are entangled if the heat current surpasses a critical heat current. Similar to our results, they found that if the population of $\ket{11}$ vanishes, a nonzero heat current becomes an entanglement witness.

\section{Conclusion and outlook}
\label{sec:conclusion}
In this paper, we introduced a continuous feedback protocol aiming to increase the stationary entanglement production of a quantum thermal machine consisting of two coherently interacting qubits, incoherently coupled to two thermal reservoirs. In the absence of feedback, it was shown in Ref.~\cite{Bohr_Brask_2015} that a temperature gradient between the reservoirs could weakly entangle the qubits. The feedback protocol only makes use of local operations, measuring the parity observable of the qubits and controlling the coupling to the warmer environment. Our investigation shows that the protocol increases the stationary entanglement production. In particular, we identified an ideal operation regime where the entanglement significantly increases. This involved putting the temperature of the colder bath to zero and performing projective measurements. In this regime, we find that the heat current across the system is proportional to the concurrence. This implies that a nonzero heat current acts as an entanglement witness. In addition, we investigated the operational usefulness of the entanglement, and found that the entanglement is capable of violating the CHSH inequality and performing quantum teleportation, which was not possible in the absence of the feedback protocol. We additionally studied the entanglement production under non-ideal conditions. Decreasing the strength of the measurement (performing nonprojective measurements) induces mistakes in the feedback, lowering the entanglement production. Similarly, the entanglement decreases when the temperature of the cold bath is nonzero, as the cold qubit can be thermally excited. 

Extensions of the protocol involves, e.g., heralding \cite{Tavakoli-Quantum-2018}, which has the potential of generating maximally entangled states, and extensions to multipartite systems. Finally, we note that several experimental platforms are available for realizing the protocol. Among these, semiconductor quantum dots and superconducting qubits are promising candidates.

\begin{acknowledgments}
We thank Patrick P.~Potts for fruitful discussions. P.S.~and B.A.A. were supported by the Swedish Research Council, Grant No.
2018-03921. P.B. is supported by grant number FQXi Grant Number:
FQXi-IAF19-07 from the Foundational Questions Institute
Fund, a donor advised fund of Silicon Valley Community
Foundation. P.B. also acknowledges funding 
from the European Research Council (Consolidator grant ’Cocoquest’ 101043705. G.F.D. acknowledges the Wallenberg Center for Quantum Technology (WACQT) for financial support via the EDU-WACQT program funded by Marianne and Marcus Wallenberg Foundation.
A.T. is supported by the Wenner-Gren Foundation and by the Knut and Alice Wallenberg Foundation through the Wallenberg Center for Quantum Technology (WACQT).
\end{acknowledgments}

\bibliographystyle{quantum}
\bibliography{refs}

\onecolumn\newpage

\hypertarget{sec:appendix}
\appendix
\section*{Appendix}
\renewcommand{\thesubsubsection}{A.\arabic{subsection}.\Roman{subsubsection}}
\renewcommand{\thesubsection}{A.\Roman{subsection}}
\renewcommand{\thesection}{A}
\numberwithin{equation}{section}
\renewcommand{\thefigure}{A.\arabic{figure}}
%

\subsection{Derivation of Eq.~(\ref{eq:MEIdealMeas})} \label{appendix: separation of time scales}
In this section, we provide a derivation for Eq.~(\ref{eq:MEIdealMeas}). The method for this derivation was first presented in Ref.~\cite{Annby-PRL-2022}. We begin by writing Eq.~(\ref{eq: Fokker-Planck Master Equation}) as
\begin{equation}
    \partial_t\r_t(D) = \mathcal{L}(D)\r_t(D) + \mathcal{F}(D)\r_t(D),
    \label{eq:compact QFPME}
\end{equation}
where we introduced the superoperator
\begin{equation}
    \mathcal{F}(D) = -\gamma\partial_D\mathcal{A}(D)+\frac{\gamma^2}{8\lambda}\partial_D^2.
\end{equation}
We are interested in the fast detector regime where $\gamma\gg\text{max}\{g,\Gamma_{kl}^{\pm}\}$. To this end, we can expand the density matrix $\r_t(D)$ in powers of $1/\gamma$. Following the procedure of Ref.~\cite{Annby-PRL-2022} (see Section II in the supplemental material of \cite{Annby-PRL-2022}), we find, to zeroth order in $1/\gamma$, that
\begin{equation}
    \r_t(D) = \sum_{aa'}\pi_{aa'}(D)\mathcal{V}_{aa'}\r_t ,
    \label{eq:D dep. density matrix}
\end{equation}
where $\r_t=\int dD\r_t(D)$ is the state of the system, $\mathcal{V}_{aa'}\r=\mel{a}{\r}{a'}\dyad{a}{a'}$, with $\ket{a}$ being the eigenstates of the observable $\hat{A}=\hat{\sigma}_z\otimes\hat{\sigma}_z$, and
\begin{equation}
    \pi_{aa'}(D) = \sqrt{\frac{4\lambda}{\pi\gamma}} e^{-\frac{4\lambda}{\gamma}[D-(\xi_a+\xi_{a'})/2]^2},
\end{equation}
with $\xi_a$ being the eigenvalues corresponding to the eigenstate $\ket{a}$. By plugging Eq.~(\ref{eq:D dep. density matrix}) into Eq.~(\ref{eq:compact QFPME}) and integrating over $D$, we find Eq.~(\ref{eq:MEIdealMeas}) with
\begin{align}
    \mathcal{L}_{\rm fb} = &\mathcal{L}\left[ (1-\eta)(\mathcal{V}_{00,00}+\mathcal{V}_{11,11})+\eta(1-\mathcal{V}_{00,00}-\mathcal{V}_{11,11})\right] + \\
    &\tilde{\mathcal{L}}\left[ \eta(\mathcal{V}_{00,00}+\mathcal{V}_{11,11})+(1-\eta)(1-\mathcal{V}_{00,00}-\mathcal{V}_{11,11})\right],
\end{align}
where we introduced the feedback error probability
\begin{equation}
    \eta = \frac{1}{2}\left[1-\erf\left( 2\sqrt{\frac{\lambda}{\gamma}}\right) \right].
\end{equation}
By vectorizing the density matrix as $\r_t=(\varrho_{00}, \varrho_{01}, \varrho_{10}, \varrho_{11}, \alpha, \alpha^*)^T$, $\mathcal{L}_{\rm fb}$ can be written in matrix form as
\tiny
\begin{align}
\label{eq:full liouvillian}
&\mathcal{L}_{\rm fb} =\left[\begin{matrix}
- \Gamma_{\s{C}0}^+ - \Gamma_{\s{H}0}^+ \left(1 - \eta\right) & \eta \Gamma_{\s{H}0}^-& \Gamma_{\s{C}0}^- & 0 & 0 & 0\\
\Gamma_{\s{H}0}^+ \left(1 - \eta\right) & - \Gamma_{\s{C}1}^+ - \eta \Gamma_{\s{H}0}^-& 0 & \Gamma_{\s{C}1}^- & i g & - i g\\
\Gamma_{\s{C}0}^+ & 0 & - \Gamma_{\s{C}0}^- & \Gamma_{\s{H}1}^- \left(1 - \eta\right) & - i g & i g\\
0 & \Gamma_{\s{C}1}^+ & \eta & - \Gamma_{\s{C}1}^- - \Gamma_{\s{H}1}^- \left(1 - \eta\right) & 0 & 0\\
0 & i g & - i g & 0 & -\frac{1}{2} \left[ \Gamma_{\s{C}1}^+ + \eta \Gamma_{\s{H}0}^- + \Gamma_{\s{C}0}^-\right] & 0\\
0 & - i g & i g & 0 & 0 & -\frac{1}{2} \left[ \Gamma_{\s{C}1}^+ + \eta \Gamma_{\s{H}0}^- + \Gamma_{\s{C}0}^-\right]
\end{matrix}\right].
\end{align}
\normalsize

\subsection{Steady state solutions to Eq.~(\ref{eq:MEIdealMeas})} \label{appendix: steady state calculations}
In this section, we provide the stationary solution to Eq.~(\ref{eq:MEIdealMeas}) for ideal operation conditions ($\lambda\to\infty$ and $T_{\rm C}\to0$) and when taking the limit $U\to\infty$. We do not present the general stationary state of Eq.~(\ref{eq:full liouvillian}) as the expressions are too long. 

The ideal operation conditions are equivalent to $\eta\to0$ (feedback always applied correctly) and $\Gamma_{\text{C}l}^+\to0$. Under these limits, Eq.~(\ref{eq:full liouvillian}) simplifies to
\begin{align}
    &\mathcal{L}_{\rm fb} =\left[\begin{matrix}
 - \Gamma_{\s{H}0}^+  & 0 & \Gamma_{\s{C}0}^- & 0 & 0 & 0\\
\Gamma_{\s{H}0}^+  & 0& 0 & \Gamma_{\s{C}1}^- & i g & - i g\\
0 & 0 &  - \Gamma_{\s{C}0}^- & \Gamma_{\s{H}1}^-  & - i g & i g\\
0 & 0 & 0 & - \Gamma_{\s{C}1}^- - \Gamma_{\s{H}1}^-  & 0 & 0\\
0 & i g & - i g & 0 & -\frac{1}{2}\Gamma_{\s{C}0}^- & 0\\
0 & - i g & i g & 0 & 0 & -\frac{1}{2}  \Gamma_{\s{C}0}^-
\end{matrix}\right]. 
\label{eq: optimal liouvillian_appendix}
\end{align}
The stationary state corresponds to the nullspace of $\mathcal{L}_{\rm fb}$, and is given by
\begin{equation}
    \r_\infty = \frac{1}{\mathcal{N}}\begin{bmatrix}
4 g^2 \Gamma_{\s{C}0}^- & 0 &0  &0 \\ 
0 & (4 g^2 + { \Gamma_{\s{C}0}^-}^2)\,  \Gamma_{\s{H}0}^+ & 2 i g \Gamma_{\s{C}0}^- \Gamma_{\s{H}0}^+ &0 \\ 
0 &-2 i g \Gamma_{\s{C}0}^- \Gamma_{\s{H}0}^+  &  4 g^2 \Gamma_{\s{H}0}^+ &0 \\ 
0 &0  &0  & 0
\end{bmatrix}, 
\label{eq:stationary-densitymatrix_fast_detector}
\end{equation}
where $\mathcal{N} = (\Gamma_{\rm C0}^{-})^2 \Gamma_{\s{H}0}^+ + 4 g^2 (\Gamma_{\s{C}0}^- + 2 \Gamma_{\s{H}0}^+)$ is a normalization constant. We note that the population of the doubly excited state vanishes, even though we did not make any assumptions on $U$. By identifying $\alpha$, $\varrho_{00}$ and $\varrho_{11}$ [Eq.~(\ref{eq: simple steady form})], and using Eq.~(\ref{eq: simple concurrence}), we get the expression for concurrence given in Eq.~(\ref{eq: ss concurrence}). We also stress that we have made no assumptions whether the baths are fermionic or bosonic, the results are valid for both.

We now look at the limit $U\to\infty$, corresponding to $\Gamma_{k1}^+\to0$. Equation (\ref{eq:full liouvillian}) simplifies to
\small
\begin{align}
\label{eq:liouvillian U infinity}
&\mathcal{L}_{\rm fb} =\left[\begin{matrix}
- \Gamma_{\s{C}0}^+ - \Gamma_{\s{H}0}^+ \left(1 - \eta\right) & \eta \Gamma_{\s{H}0}^-& \Gamma_{\s{C}0}^- & 0 & 0 & 0\\
\Gamma_{\s{H}0}^+ \left(1 - \eta\right) & - \eta \Gamma_{\s{H}0}^-& 0 & \Gamma_{\s{C}1}^- & i g & - i g\\
\Gamma_{\s{C}0}^+ & 0 & - \Gamma_{\s{C}0}^- & \Gamma_{\s{H}1}^- \left(1 - \eta\right) & - i g & i g\\
0 & 0 & 0 & - \Gamma_{\s{C}1}^- - \Gamma_{\s{H}1}^- \left(1 - \eta\right) & 0 & 0\\
0 & i g & - i g & 0 & -\frac{1}{2} [ \eta \,  \Gamma_{\s{H}0}^- + \Gamma_{\s{C}0}^-] & 0\\
0 & - i g & i g & 0 & 0 & -\frac{1}{2} [\eta  \Gamma_{\s{H}0}^- + \Gamma_{\s{C}0}^-]
\end{matrix}\right].
\end{align}
\normalsize
Note that all rates accompanied by $1-\eta$ correspond to transitions when applying feedback correctly, and all rates with an $\eta$ correspond to transitions when feedback is applied wrongly. Similar to the ideal operation conditions, $\ket{11}$ is decoupled from the remaining states in the stationary limit. Here we write the stationary state as
\begin{align}
        \r_\infty = \frac{1}{\tilde{\mathcal{N}}} \begin{bmatrix}
\varrho_{00}' &0  & 0 &0 \\ 
 0& \varrho_{01}' & \alpha' &0 \\ 
 0& {\alpha'}^* & \varrho_{10}' &0 \\ 
 0& 0 & 0 & \varrho_{11}'
\end{bmatrix},
\end{align}
where $\tilde{\mathcal{N}}$ is a normalization constant. The matrix elements and normalization constant are given by
\begin{align}
    \varrho_{00}' &= (\Gamma _{\text{C0}}^-+\eta  \Gamma _{\text{H0}}^-) \,(\eta  \Gamma _{\text{C0}}^- \Gamma _{\text{H0}}^-+4 g^2), \\
    \varrho_{01}' &= 4 g^2 \Gamma _{\text{C0}}^++(1-\eta) \Gamma _{\text{H0}}^+\, [{\Gamma _{\text{C0}}^-}^2+\eta \Gamma _{\text{C0}}^- \Gamma _{\text{H0}}^-+4 g^2], \\
    \varrho_{10}' &= 4 (1-\eta) g^2 \Gamma _{\text{H0}}^+ + \,\Gamma _{\text{C0}}^+\, [\eta \Gamma _{\text{H0}}^- \,(\Gamma _{\text{C0}}^-+\eta  \Gamma _{\text{H0}}^-)+4 g^2], \\
    \varrho_{11}' &= 0, \\
    \alpha' &= 2 i g \,[(1-\eta ) \Gamma _{\text{C0}}^- \Gamma _{\text{H0}}^+-\eta  \Gamma _{\text{C0}}^+ \Gamma _{\text{H0}}^-], \\
    \tilde{\mathcal{N}} &= 8 g^2 \,[\Gamma _{\text{C0}}^++(1-\eta) \Gamma _{\text{H0}}^+]+\Gamma _{\text{C0}}^- \,\left[\,\eta \Gamma _{\text{H0}}^- \,[\Gamma _{\text{C0}}^++\eta  \Gamma _{\text{H0}}^-+(1-\eta) \Gamma _{\text{H0}}^+]\,+\,4 g^2\right] \\ &+\eta ^2 \Gamma _{\text{C0}}^+ {\Gamma _{\text{H0}}^-}^2+\,{\Gamma _{\text{C0}}^-}^2 \, [\eta  \Gamma _{\text{H0}}^-+(1-\eta) \Gamma _{\text{H0}}^+]+\,4 \eta  g^2 \Gamma _{\text{H0}}^-.
\end{align}

\subsection{Optimizing teleportation fidelity} \label{appendix: singlet fraction}
In this section, we show that the teleporation fidelity is maximized for the same parameters that maximizes the concurrence in Sec.~\ref{sec: results}. To do so, we work with the optimal operation conditions $\lambda \to \infty$ and $\tc \to 0$. To carry out an optimization for the teleportation fidelity, we follow a standard optimization scheme and consider the gradient $\nabla \equiv ( \partial_{g}, \partial_{\gmmc}, \partial_{\gmmh})^{\text{T}}$ of Eq.~(\ref{eq: Singlet Fraction}) for a general steady state of the Liouvillian in Eq.~(\ref{eq: optimal liouvillian}). Doing so leads us to the expressions

\begin{align}
\nabla F(\r_\infty) =  
\begin{cases}
    \begin{matrix}
\mathcal{N}\begin{pmatrix}
2\gmmc\gmmh\,[ -2\gmmc g\,(\gmmc + 2g ) + \,( {\gmmc}^2-8g^2 )\,\gmmh]\\ 
2 g \gmmh \,({\gmmc}^2-8 g^2)\, (g-\gmmh)\\ 
2 g^2\gmmc  \,({\gmmc}^2+4 g\gmmc +8 g^2)
\end{pmatrix} & \hspace{0.5cm}\s{if} \hspace{0.1cm} 1+2\alpha - 2\Delta \leq 0 \\
-\mathcal{N}\begin{pmatrix}
-4 g{\gmmc}^3\gmmh\\ 
2 g^2 \gmmh \,({\gmmc}^2-8 g^2)\\
2 g^2 \gmmc  \,({\gmmc}^2+8
   g^2)
\end{pmatrix} \,\,\,\,\,\,\,\,\,\,\,& \hspace{0.5cm}\s{otherwise}
\end{matrix}
\end{cases}
\label{appendix: singlet f. gradient 2}
\end{align}
where $\mathcal{N} = ({\gmmc}^2 \gmmh+4 g^2 \gmmc +8 g^2 \gmmh)^{-2}$. Using Eqs.~(\ref{appendix: singlet f. gradient 2}) we find that regardless of whether or not $1+2\alpha - \Delta \leq 0$, optimal fidelity is obtained for $g = {\gmmc}/{2\sqrt{2}}$, which is the same result obtained when optimizing the system concurrence. By inserting this value of $g$ in Eq.~(\ref{eq: Singlet Fraction}), we obtain 

\begin{align}
    \mathcal{F}(\r_\infty) = \frac{1}{3}\,\left({1 + \,\frac{(4 + 2\sqrt{2}) \gmmh}{\gmmc + 4 \gmmh } }\right),
\end{align}
which is a monotonic function in $\gh^+$. Thus, considering $\gmmh \gg \gmmc$ we obtain the optimal solution $\mathcal{F}(\r_\infty) = (4+\sqrt{2})/6\approx0.90$ that we presented in Sec.~\ref{sec: results}.

\subsection{Optimizing CHSH}
\label{appendix:CHSH}
In this section, we show that the CHSH is maximized for the same parameters that maximizes the concurrence in Sec.~\ref{sec: results}. As in the case with fidelity, we carry out this optimization procedure by analytically obtaining an expression for the gradient $\nabla \equiv ( \partial_g, \partial_{\gmmc}, \partial_{\gmmh})$ of Eq.~(\ref{eq: CHSH}) for ideal operation conditions. This is carried out by separating the cases in which $4\alpha ^2 - (2\Delta - 1 ) ^2 \leq 0$ from all others. In the case where  $4\alpha ^2 - (2\Delta - 1 ) ^2 > 0$, the gradient of Eq.~(\ref{eq: CHSH})  evaluated for the steady state of Eq.~(\ref{eq: Liouvillian optimal conditions}) returns
\begin{align}
    \nabla \text{CHSH} = \begin{pmatrix}
g {\gmmh}^2 {\gmmc}^2\left(64\sqrt{2} g^2-8\sqrt{2} {\gmmc}^2\right) \\ 
g^2 \gmmh \gmmc\left(8\sqrt{2}{\gmmc}^2 \gmmh-32\sqrt{2} \gmmc g^2-64\sqrt{2} g^2 \gmmh\right)\\
32\sqrt{2} \left(\frac{{\gmmc}^4 g^3{\gmmh}^6}{ \sqrt{{\gmmc}^2 \gmmh+4 \gmmc g^2+8 g^2 \gmmh}}\right)
\end{pmatrix}. \label{eq: chsh gradient 1}
\end{align}
Setting Eq.~(\ref{eq: chsh gradient 1}) to zero and solving for $g$, $\gmmc$ and $\gmmh$ returns a maximum value of $\s{CHSH} = 2$ at $g = \frac{\gmmc \sqrt{\gmmh}}{\sqrt{4\gmmc +8\gmmh}}$. However, if we instead carry out the optimization procedure, by considering the gradient of Eq.~(\ref{eq: CHSH})  for $4\alpha ^2 - (2\Delta - 1 ) ^2 \leq 0$, we obtain 
\begin{align}\nabla \text{CHSH} = \begin{pmatrix}
\frac{g^2 \gmmh \left({\gmmc}^3 \left(-64 g^2-32 \gh^{+^2}\right)+16 {\gmmc}^4 \gmmh+\gmmc \left(256 g^2 {\gmmh}^2+512 g^4\right)-1024 g^4 \gmmh\right)}{\left( {\gmmc}^2 \gmmh+4 \gmmc g^2+8 g^2 \gmmh\right)^2 \sqrt{{\gmmc}^2
   \left(32 g^2 {\gmmh}^2+16 g^4\right)-8 {\gmmc}^3 g^2 \gmmh+ {\gmmc}^4 {\gmmh}^2-64 \gmmc g^4 \gmmh+64 g^4 {\gmmh}^2}}\\
\frac{{\gmmc}^2 g \gmmh \left({\gmmc}^2 \left(128 g^2+32
   {\gmmh}^2\right)-32 {\gmmc}^3 \gmmh-384 \gmmc g^2 \gmmh-256 g^2 {\gmmh}^2\right)}{\left( {\gmmc}^2 \gmmh+4 \gmmc g^2+8 g^2 \gmmh\right)^2 \sqrt{{\gmmc}^2 \left(32 g^2 {\gmmh}^2+16 g^4\right)-8 {\gmmc}^3 g^2 \gmmh+ {\gmmc}^4 {\gmmh}^2-64 \gmmc g^4
   \gmmh+64 g^4 {\gmmh}^2}}\\
\frac{\gmmc g^2 \left(384 {\gmmc}^2 g^2 \gmmh-64 {\gmmc}^3 g^2+16 {\gmmc}^4 \gmmh-512 \gmmc g^4+1024 g^4 \gmmh\right)}{\left( {\gmmc}^2 \gmmh+4 \gmmc g^2+8 g^2
   \gmmh\right)^2 \sqrt{{\gmmc}^2 \left(32 g^2 {\gmmh}^2+16 g^4\right)-8 {\gmmc}^3 g^2 \gmmh+ {\gmmc}^4 {\gmmh}^2-64 \gmmc g^4 \gmmh+64 g^4 {\gmmh}^2}}
\end{pmatrix}, \label{eq: chsh gradient 2}
\end{align}
which has optimal solution at  $g = \frac{\gmmc}{2\sqrt{2}}$ and $\gmmh \gg \gmmc$, where $\text{CHSH} = \sqrt{6} \approx 2.45$.

\subsection{Heat current}
\label{app:heat exchange}
Here we calculate the stationary heat current in the system. To this end, it is useful to decompose the feedback Liouvillian as $\mathcal{L}(D)=\theta(D)(\mathcal{L}_{\rm C}+\mathcal{L}_{\rm H}) + [1-\theta(D)]\mathcal{L}_{\rm C}$, where
\begin{equation}
    \mathcal{L}_{k} = \Gamma_{k0}^+\mathcal{D}[\hat{J}_{k0}^\dagger] + \Gamma_{k1}^+\mathcal{D}[\hat{J}_{k1}^\dagger] +
    \Gamma_{k0}^-\mathcal{D}[\hat{J}_{k0}] +
    \Gamma_{k1}^-\mathcal{D}[\hat{J}_{k1}], \hspace{1cm} k=\rm{C,H}.
\end{equation}
The average energy of the system is given by $E=\int dD\trace\{\hat{H}\r_t(D)\}$. Taking the time derivative of this gives
\begin{align}
    \dot{E}&=-\int_{-\infty}^\infty dD\trace\{\hat{H}\partial_t\r_t(D)\} \\
    &= -\int_{-\infty}^\infty dD\trace\{\hat{H}\mathcal{L}(D)\r_t(D)\} \\
    &= \trace\{\hat{H}\mathcal{L}_{\rm C}\r_t\} + \int_{0}^\infty dD\trace\{\hat{H}\mathcal{L}_{\rm H}\r_t(D)\} \\
    &= \dot{Q}_{\rm C} + \dot{Q}_{\rm H},
\end{align}
where the minus sign decides the sign convention, i.e., a positive sign corresponds to heat flowing into the baths, and we used that $\r_t=\int_{-\infty}^\infty dD\r_t(D)$, $\int_{-\infty}^\infty dD \left[ -\gamma\partial_D\mathcal{A}(D)+\frac{\gamma^2}{8\lambda}\partial_D^2 \right]\r_t(D)=0$, and we introduced the following heat currents associated with the cold (C) and hot (H) baths, 
\begin{equation}
    \dot{Q}_{\rm C} = -\trace\{\hat{H}\mathcal{L}_{\rm C}\r_t\} \hspace{1cm} \text{and} \hspace{1cm} \dot{Q}_{\rm H} = -\int_{0}^\infty dD\trace\{\hat{H}\mathcal{L}_{\rm H}\r_t(D)\}.
\end{equation}
In steady state, we get the heat current $\dot{Q}=\dot{Q}_{\rm C}=-\dot{Q}_{\rm H}$. By using $\dot{Q}_{\rm C}$, we get
\begin{equation}
    \dot{Q} = (\varepsilon+U)\Gamma_{\rm C1}^-\varrho_{11} + \varepsilon\Gamma_{\rm C0}^-\varrho_{10} - \varepsilon \Gamma_{\rm C0}^+ \varrho_{00} - (\varepsilon+U)\Gamma_{\rm C1}^{+}\varrho_{01}.
    \label{eq: general heat current}
\end{equation}
Taking the limits $\tc\to0$ ($\Gamma_{\text{C}l}^+\to0$) and $\lambda\to\infty$, we get, for $\gamma\gg\text{max}\{g,\Gamma_{kl}^\pm\}$, that
\begin{align}
    \dot{Q} = \varepsilon\Gamma_{\text{C}0}^-\varrho_{10} = - \Gamma_{\rm C0}^-\trace\{\hat{H}\mathcal{D}[\hat{J}_{\rm C0}]\r_\infty\}, 
\end{align}
as specified in Eq.~(\ref{eq: concurrence as funct of heat}) in the main text. By explicit calculation, using the stationary state in Eq.~(\ref{eq:stationary-densitymatrix_fast_detector}), we obtain the relation $\dot{Q}=\varepsilon g\mathcal{C}$ as specified in the main text.

We may also investigate the heat current for $U\to\infty$ ($\Gamma_{k1}^+\to0$). Under this limit, Eq.~(\ref{eq: general heat current}) is simplified to
\begin{equation}
    \dot{Q} = \varepsilon ( \Gamma_{\rm C0}^-\varrho_{10} - \Gamma_{\rm C0}^+ \varrho_{00} ).
\end{equation}
Using the expressions from Appendix \ref{appendix: steady state calculations}, we get
\begin{equation}
    |\dot{Q}| = \frac{ 4 g^2 \varepsilon  |(1-\eta) \Gamma _{\text{C0}}^- \Gamma _{\text{H0}}^+ - \eta \Gamma _{\text{C0}}^+ \Gamma _{\text{H0}}^-| }{\tilde{\mathcal{N}}} = 2\varepsilon g \frac{|\alpha'|}{\Tilde{\mathcal{N}}} = \varepsilon g \mathcal{C},
\end{equation}
where we used Eq.~(\ref{eq: simple concurrence}). This proves that the heat current is proportional to the concurrence when $U\to\infty$.

\subsection{Concurrence, CHSH, and teleportation fidelity beyond ideal operation conditions}
\label{appendix:concurrence,CHSH,fidelity}
In Sec.~\ref{beyond ideal operation} we studied the steady state concurrence of the system outside the regime of ideal operation conditions, see Sec.~\ref{sec: results}. In this section, we study CHSH and the teleportation fidelity for both bosonic and fermionic reservoirs beyond the ideal operation conditions. For the bosons, we only consider $U=0$, as the reservoirs naturally would consist of photons or phonons, and thereby, do not carry charge.

\begin{figure}[H] 
\centering
 \makebox[\textwidth]{\includegraphics[width=.8\paperwidth]{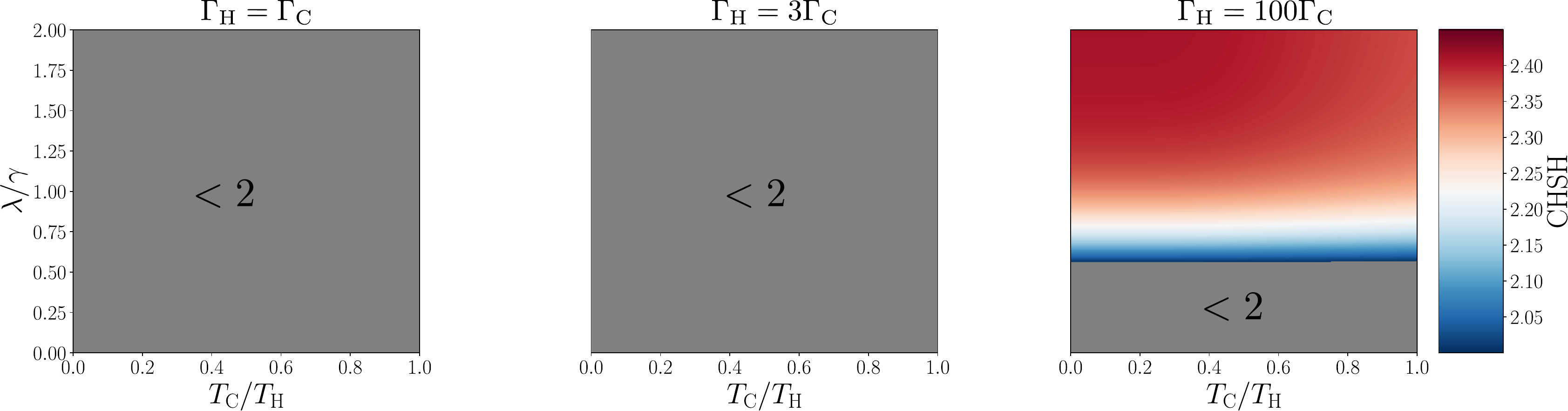}}
\caption{Statiornay CHSH for fermionic reservoirs as a function of the measurement strength $\lambda$ and temperature of the cold reservoir $\tc$. Each heat map was then plotted for a different choice of the ratio $\gh / \gc$. The three graphs were obtained considering $\gc/\varepsilon = 10^{-3}$, $U/\varepsilon = 100;$  $g = \frac{\gc}{2\sqrt{2}}$; $\gamma = 1$ and $\th/\varepsilon = 1$.}
\label{fig: finite lambda CHSH fermionic gh}
\end{figure}

\begin{figure}[H] 
\centering
 \makebox[\textwidth]{\includegraphics[width=.8\paperwidth]{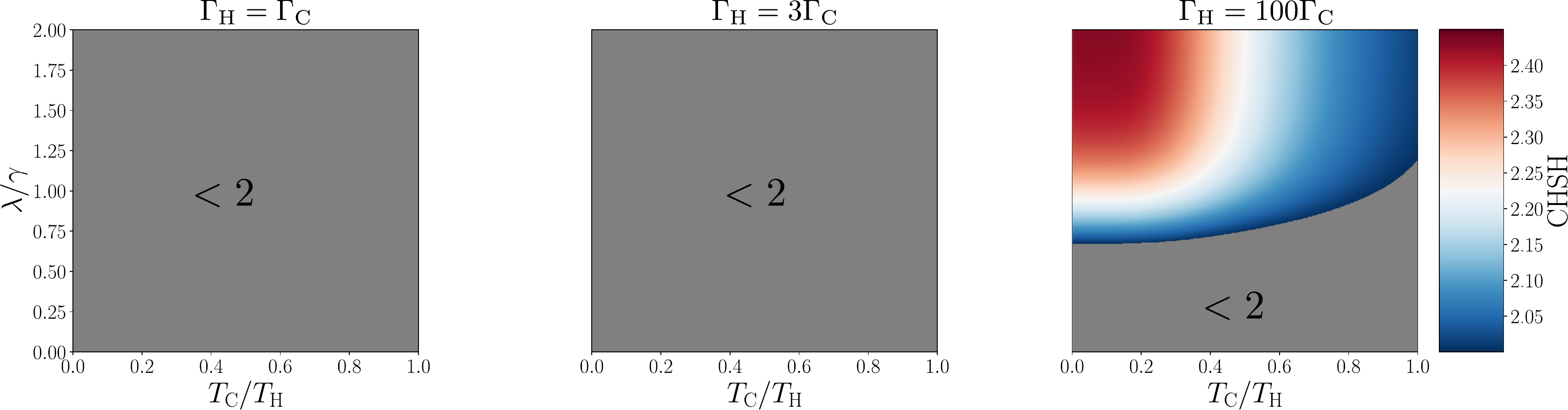}}
\caption{Stationary CHSH for bosonic reservoirs as a function of the measurement strength $\lambda$ and temperature of the cold reservoir $\tc$. Each heat map was then plotted for a different choice of the ratio $\gh / \gc$.The three graphs were obtained considering $\gc/\varepsilon = 10^{-3}$, $U = 0$,  $g = \frac{\gc}{2\sqrt{2}}$; $\gamma = 1$ and $\th/\varepsilon = 1$.}
\label{fig: finite lambda CHSH bosonic gh}
\end{figure}

Figure~\ref{fig: finite lambda CHSH fermionic gh} shows CHSH of the fermionic system as a function of $\lambda$ and $\tc$ for different choices of $\gh$. Similarly to what we found in Sec.~\ref{sec: results}, from Fig.~\ref{fig: finite lambda CHSH fermionic gh} we see that increasing the coupling to the hot reservoir results in higher values of CHSH. However, a violation of Bell's inequality is achieved only in the limit $\gh \gg \gc$. On the other hand, if we compare the results obtained for the fermionic system with the ones obtained for the bosonic engine (see Fig.~\ref{fig: finite lambda CHSH bosonic gh}), we notice that within similar parameter regimes the bosonic engine is capable of producing higher CHSH values. This is due to the difference between the Bose-Einstein and Fermi-Dirac distributions, which allows the bosonic engine to produce higher heat currents, see Eq.~(\ref{eq: rates}). However, a stronger coupling to the hot reservoir is still required in order to violate Bell's inequality.  

\begin{figure}[H] 
\centering
 \makebox[\textwidth]{\includegraphics[width=.8\paperwidth]{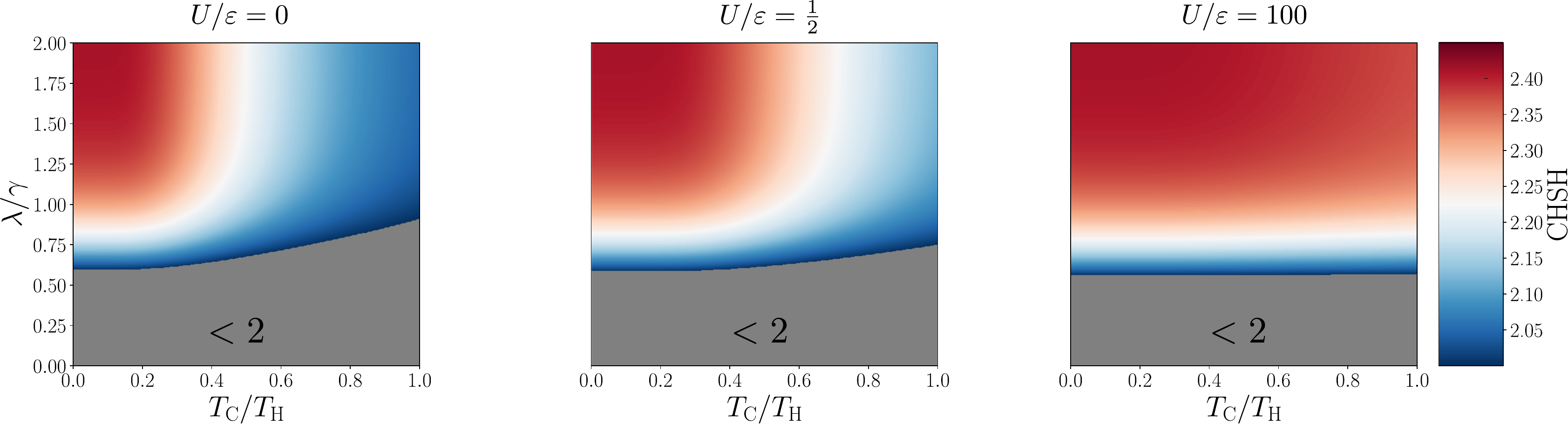}}
\caption{Stationary CHSH for fermionic reservoirs as a function of the measurement strength $\lambda$ and temperature of the cold reservoir $\tc$. Each heat map was then plotted for a different choices of $U$. The three graphs were obtained considering $\gc/\varepsilon = 10^{-3}$, $\gh = 100\gc;$  $g = \frac{\gc}{2\sqrt{2}}$; $\gamma = 1$ and $\th/\varepsilon = 1$.}
\label{fig: finite lambda CHSH fermionic U}
\end{figure}

In Fig.~\ref{fig: finite lambda CHSH fermionic U} we instead study the CHSH for the fermionic engine as a function of $\lambda$ and $\tc$ for different choices of interqubit interaction strength $U$. From these maps we see that for $U/\varepsilon=0$ we recover the optimal CHSH value when within the limit $\lambda \to \infty$ and $\tc \to 0$. Furthermore, increasing the value of $U$ allows for the region of optimal CHSH value to extend to higher values of $\tc$, provided a high enough measurement strength $\lambda$ (note $\gh \gg \gc$).  

\begin{figure}[H] 
\centering
 \makebox[\textwidth]{\includegraphics[width=.8\paperwidth]{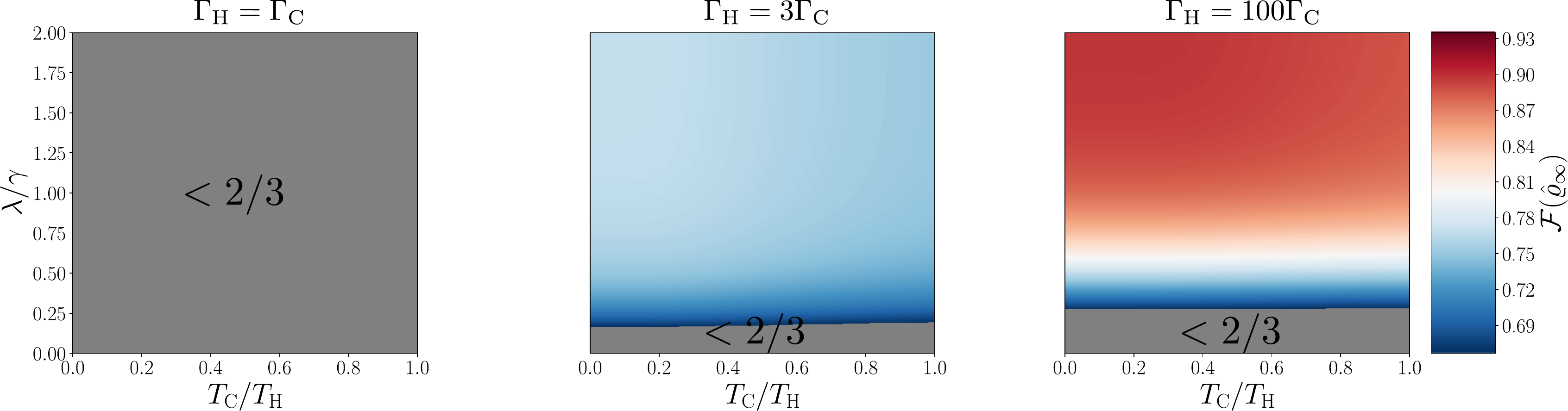}}
\caption{Teleportation fidelity for fermionic reservoirs as a function of the measurement strength $\lambda$ and temperature of the cold reservoir $\tc$. Each heat map was then plotted for different choices of $\gh$. The three graphs were obtained considering $\gc/\varepsilon = 10^{-3}$, $U/\varepsilon =100;$  $g = \frac{\gc}{2\sqrt{2}}$; $\gamma = 1$ and $\th/\varepsilon = 1$.}
\label{fig: finite lambda SFRACT fermionic gh}
\end{figure}

\begin{figure}[H] 
\centering
 \makebox[\textwidth]{\includegraphics[width=.8\paperwidth]{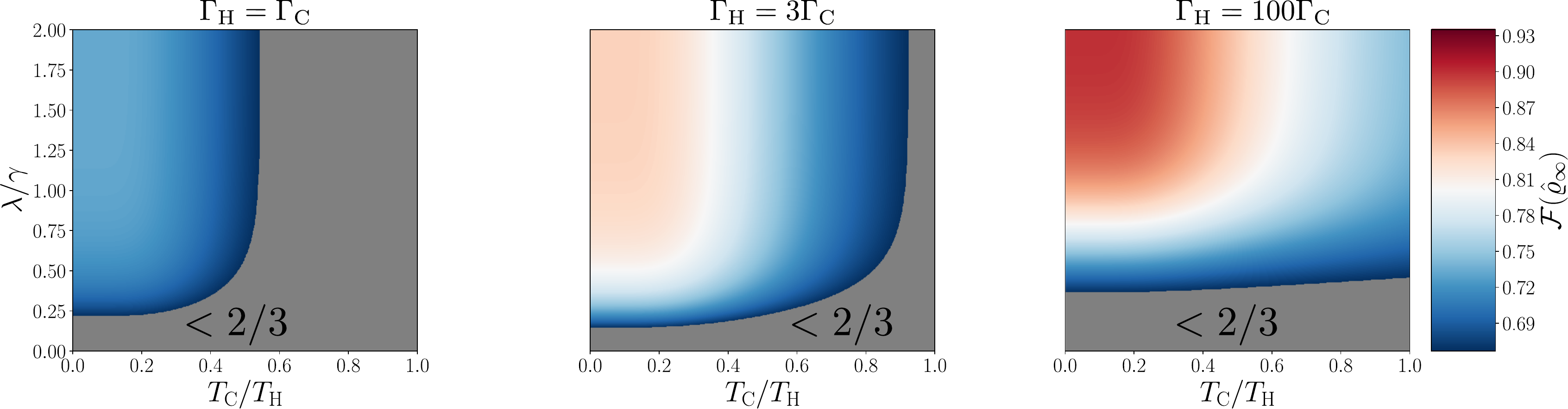}}
\caption{Teleportation fidelity for bosonic reservoirs as a function of the measurement strength $\lambda$ and temperature of the cold reservoir $\tc$. Each heat map was then plotted for different choices of $\gh$. The three graphs were obtained considering $\gc/\varepsilon = 10^{-3}$, $U = 0$,  $g = \frac{\gc}{2\sqrt{2}}$; $\gamma = 1$ and $\th/\varepsilon = 1$.}
\label{fig: finite lambda SFRACT bosonic gh}
\end{figure}

Figure~\ref{fig: finite lambda SFRACT fermionic gh} shows the teleportation fidelity of the fermionic engine as a function of $\lambda$ and $\tc$ for different choices of $\gh$. When comparing these plots with the ones for CHSH in Fig.~\ref{fig: finite lambda CHSH fermionic gh}, we find that, in this scenario, teleportation fidelity represents a weaker condition for the usefulness of the entanglement contained in the engine, as the entanglement can uphold teleportation already for $\gh = 3\gc$, whilst, according to CHSH, we required $\gh \gg \gc$. Furthermore, we see that this is yet more apparent in the case of the bosonic engine (see Fig.~\ref{fig: finite lambda SFRACT bosonic gh}), where the entanglement produced by the engine is operationally useful already at $\gh = \gc$.

\begin{figure}[H] 
\centering
 \makebox[\textwidth]{\includegraphics[width=.8\paperwidth]{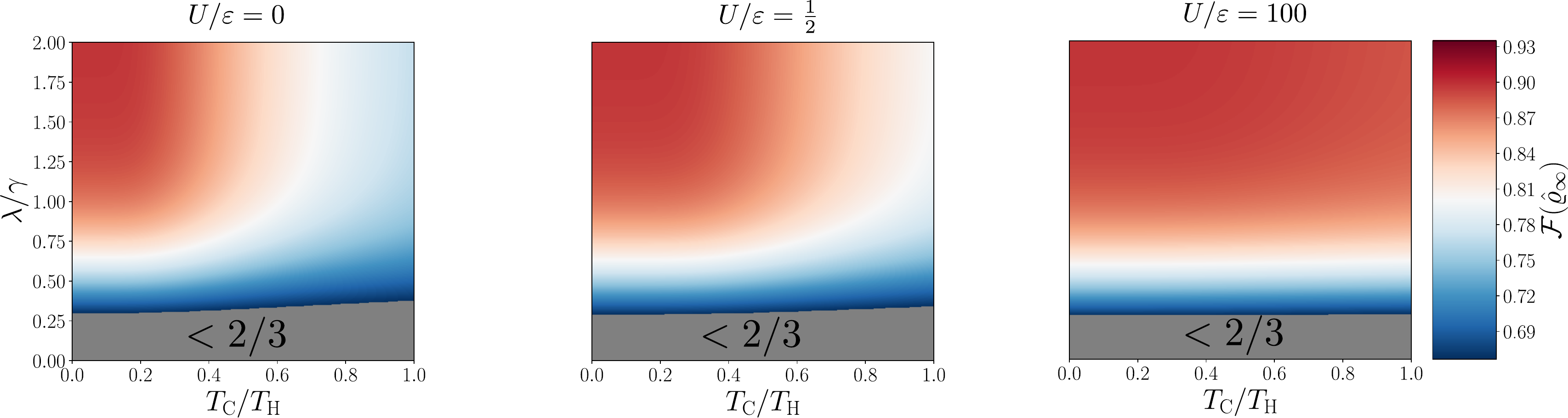}}
\caption{Teleportation fidelity for fermionic reservoirs as a function of the measurement strength $\lambda$ and temperature of the cold reservoir $\tc$. Each heat map was then plotted for different choices of $U$. The three graphs were obtained considering $\gc/\varepsilon = 10^{-3}$, $\gh = 100\gc;$  $g = \frac{\gc}{2\sqrt{2}}$; $\gamma = 1$ and $\th/\varepsilon = 1$.}
\label{fig: finite lambda SFRACT fermionic U}
\end{figure}

In Fig.~\ref{fig: finite lambda SFRACT fermionic U} we show the steady state teleportation fidelity for the fermionic engine as a function of $\lambda$ and $\tc$, for different choices of $U$. As in the case of CHSH (see Fig.~\ref{fig: finite lambda CHSH fermionic U}) we see that increasing $U$ allows the engine to reproduce the optimal fidelity results (see Sec.~\ref{sec: results}) at higher temperatures for the cold bath, granted a high enough measurement strength $\lambda$.

\begin{figure}[H] 
\centering
 \makebox[\textwidth]{\includegraphics[width=.8\paperwidth]{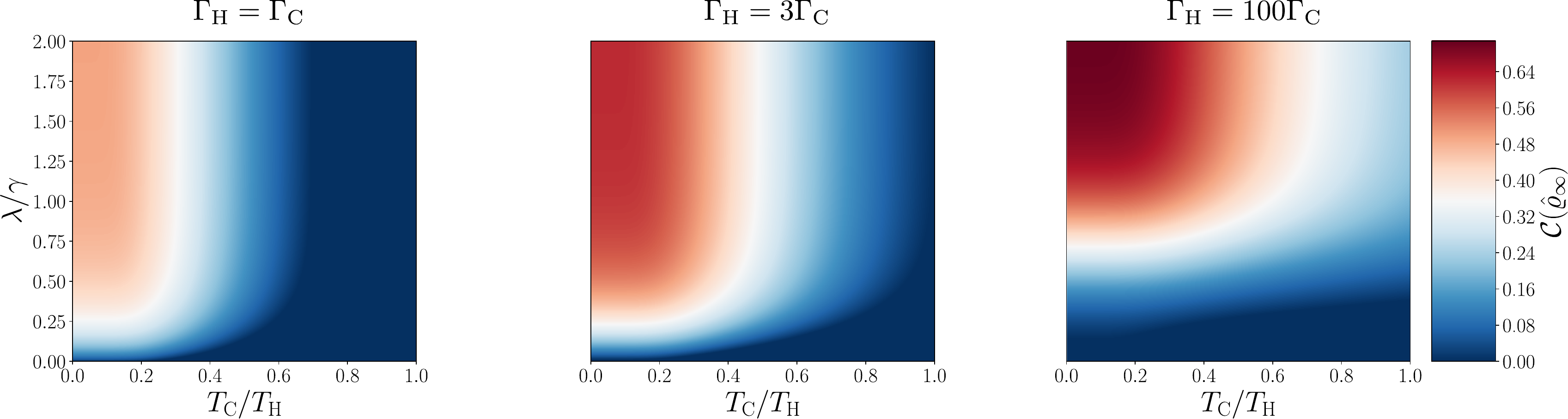}}
\caption{Steady state concurrence for bosonic reservoirs as a function of $\lambda$ and $\tc$. The three graphs were obtained considering $\gc/\varepsilon = 10^{-3}$, $U = 0$,  $g = \frac{\gc}{2\sqrt{2}}$; $\gamma = 1$ and $\th/\varepsilon = 1$.}
\label{fig: finite lambda Concurrence Bosonic gh}
\end{figure}

Figure \ref{fig: finite lambda Concurrence Bosonic gh}, shows the stationary concurrence for bosonic reservoirs as a function of $\lambda$ and $T_{\rm C}$. The results are similar to those in Fig.~\ref{fig: finite lambda different gh}, where we studied fermionic reservoirs.

\end{document}